\documentclass[prd,aps,amsfonts,preprint,floats,tightenlines,floatfix,
nofootinbib,showpacs,showkeys]{revtex4}
\usepackage{exscale}                  
\usepackage[intlimits]{amsmath}       
\usepackage{amsfonts}
\usepackage{amssymb,amscd}
\usepackage{epsfig}
\usepackage{wrapfig}

\def\dbar{{\mathchar'26\mkern-11mu d}}
\def\ellbar{{\mathchar'26\mkern-10mu \ell}}

\newcommand{\cO}{{\cal O}}
\newcommand{\cJ}{{\cal J}}

\newcommand{\ing}{\includegraphics}

\newcommand{\nn}{\nonumber}

\newcommand{\cN}{{\cal N}}

\newcommand{\cD}{{\cal D}}
\newcommand{\cL}{{\cal L}}

\newcommand{\cZ}{{\cal Z}}
\newcommand{\bi}{\bigskip}
\newcommand{\no}{\noindent}
\newcommand{\bee}{\begin{eqnarray}}
\newcommand{\eeq}{\end{eqnarray}}
\newcommand{\hk}{\hspace{0.1cm}}

\def\ellslash{\ell\kern-.5em\slash}

\newcommand{\e}{\text{e}}

\newcommand{\lla}{\left\langle}
\newcommand{\rra}{\right\rangle}

\newcommand{\rarr}{\rightarrow}

\begin{document}

\date{\today}
\title{


Infrared analysis of propagators and vertices of Yang--Mills theory in Landau
and Coulomb gauge
}

\author{Wolfgang Schleifenbaum}
\affiliation{Institute of Theoretical Physics, U.\ of T\"ubingen, 
D-72076 T\"ubingen, Germany}

\author{Markus Leder}
\affiliation{Institute of Theoretical Physics, U.\ of T\"ubingen, 
D-72076 T\"ubingen, Germany}

\author{Hugo Reinhardt}
\affiliation{Institute of Theoretical Physics, U.\ of T\"ubingen, 
D-72076 T\"ubingen, Germany}

\begin{abstract}
The infrared behaviour of gluon and ghost propagators, ghost-gluon vertex and
three-gluon vertex is investigated for both the covariant Landau and
the non-covariant Coulomb gauge. Assuming infrared ghost dominance, we find a unique infrared exponent in the $d = 4$ Landau gauge, while in the $d = 3 + 1$ Coulomb
gauge we find two different infrared exponents. 
We also show that a finite dressing of the ghost-gluon vertex has no influence on the
infrared exponents. Finally, we determine the infrared behaviour of the three-gluon vertex analytically and calculate it numerically at the symmetric point in the Coulomb gauge.
\end{abstract} 

\maketitle

\section{Introduction}
\bi

\no
In recent years there have been extensive non-perturbative studies of continuum
Yang-Mills theory using Dyson--Schwinger equations in covariant Landau gauge,
ref.\ \cite{AlkSme00,AlkFisLla04} and in the canonical quantisation approach in the Coulomb gauge, ref.\ \cite{ChrLee80}. In the latter Gaussian types of wave functionals have been used for a
variational solution of the Yang-Mills Schr\"odinger equation for the vacuum \cite{FeuRei04,ReiFeu04,Szc}.
Minimisation of the vacuum energy density gives rise to Dyson--Schwinger
equations, which are very similar to the ones arising in the functional integral
approach in covariant Landau gauge. Some of the relevant Green functions have
also been calculated on the lattice in Landau gauge in refs.\ \cite{LanReiGat01,CucMenMih04,Ste+05} and in the Coulomb gauge in ref.\ \cite{LanMoy04}. The Green functions obtained by
solving the Dyson--Schwinger equations to 1-loop order, are qualitatively very
similar to the ones obtained in the lattice calculations, at least in the case of the Landau gauge.
\bi

\no
In the present paper we are interested in the infrared limit of the basic
propagators and vertices arising in the canonical quantisation approach of
Yang-Mills theory in the Coulomb gauge. With an appropriate choice of the vacuum wave
functional, the corresponding generating functional of the Green functions is
structurally very similar to the one of the functional integral approach to
Yang-Mills theory in Landau gauge, differing only in the number of relevant
dimensions and in the precise form of the action. In both cases the infrared
limit of the generating functional is governed by the ghost sector, which (up to
the number of dimensions) is the same in both gauges. Therefore we can treat
both approaches simultaneously. Throughout the paper, Landau gauge will refer to the path integral quantisation approach whereas Coulomb gauge will refer to the canonical quantisation approach.
\bi

\no
Previously, the infrared limit of the gluon and ghost propagators in Landau and
Coulomb gauges, were investigated in ref.\ \cite{Zwa02,LerSme02} and in
ref.\ \cite{FeuRei04}, respectively. In ref.\ \cite{Zwa02} two
different solutions for the infrared exponents in Landau gauge were found, while in ref.\ \cite{FeuRei04} using the angular approximation only one solution for the infrared exponents was found (in the canonical approach) in the Coulomb gauge. We
also note, that in ref.\ \cite{Sch+05} the infrared behaviour of the ghost-gluon vertex in the Landau gauge has been studied.
\bi

\no
In this paper we perform a thorough infrared analysis of the ghost and gluon
propagators as well as of the ghost-gluon and three-gluon vertices for both Landau and Coulomb gauge without resorting to the angular approximation. We discuss the validity of the previously known solutions for the infrared exponents of the propagators. The role of the ghost-gluon vertex in loop integrals is investigated and certain infrared limits of this vertex are explicitly calculated. Analytically, the infrared behaviour of the three-gluon vertex is determined at the symmetric point and for one vanishing external momentum. Numerically, we calculate the three-gluon vertex in the Coulomb gauge at the symmetric point over the whole momentum range. Given the Green functions in the infrared, we can confirm the self-consistency of these solutions. Also, the infrared fixed point of the running coupling for both the Coulomb and Landau gauge is calculated from the infrared behaviour of the Green functions considered.
\bi

\no
The paper is organised as follows: In the section \ref{formalism} we present the infrared
form of the generating functional of the Green function in Coulomb and Landau
gauges. In section \ref{sec:analyt} we perform the infrared analysis of the various Green
functions: the ghost and gluon propagators, the ghost-gluon vertex and finally
the three-gluon vertex. Here, we also discuss the running coupling constant in the infrared. 
A short summary
and our conclusions are given in section \ref{summary}. Some mathematical details of our
infrared analysis  of the Green functions are presented in two appendices.
\bi

\no
\section{The generating functional for the infrared}
\label{formalism}
\bi

\no
The generating functional of the Green functions of Euclidean Yang-Mills theory defined
by a Lagrange density $\cL_{YM}$ is given by
\bee
\label{ZLandau}
Z[j] = {\cal N}^2 \int \mathcal{D} A \, \cJ[A]\exp \left[ - \int d^4 x \: \cL_{YM}(x) +  \int
d^4 x j^a_\mu ({x}) A^a_\mu ({x}) \right] \hk \: ,
\eeq
where the functional integration is over gauge fields, which are restricted by a
gauge condition and $J [A]$ denotes the corresponding Faddeev-Popov determinant.
A common and for perturbation theory convenient gauge is the covariant Landau
gauge
\bee
\label{2}
\partial \cdot A = 0 \hk ,
\eeq
in which the Faddeev-Popov determinant is given by 
\bee
\label{3}
J [A] = \det (- \partial \cdot D [A])  \hk .
\eeq
Here, $D [A] = \partial + g A$ denotes the covariant derivative in the adjoint
representation of the gauge group, i.e.\ $A^{ab}=A^cf^{acb}$ with $f^{abc}$ being the structure constants.
\bi

\no
The generating functional (\ref{ZLandau}) in the Landau gauge (\ref{2}) is also the
starting point for the derivation of the coupled set of Dyson-Schwinger
equations, which to 1-loop level have been extensively investigated in recent
years. For a review see ref.\ \cite{AlkSme00}. 
\bi

\no
The Faddeev-Popov determinant represents the Jacobian of the transformation to
the (curvilinear) transverse ``coordinates'' $A^\perp$, satisfying the gauge
condition (\ref{2}). The appearance of $J [A]$ turns out to have crucial
physical consequences. In order to pick a single gauge field out of the gauge
orbit $A^U = \frac{1}{g} U D U^\dagger, U \in SU (N_c)$, 
it is necessary to choose configurations
from within the (first) Gribov region $\Omega$ or more precisely from the
so-called fundamental modular region, which is a compact subset of the Gribov
region and free of gauge copies. 
\bi

\no
In the canonical quantisation approach to Yang-Mills theory, one uses the Weyl
gauge $A_0 = 0$ to avoid the problems arising from a vanishing of the canonical
momentum conjugate to $A_0$. Furthermore, Gauss' law, which here is a constraint on the wave functional to guarantee gauge
invariance, is conveniently resolved in the Coulomb gauge defined by Eq.\ (\ref{2}) for $A_0 = 0$ \cite{ChrLee80}. The Yang-Mills Schr\"odinger equation in the Coulomb gauge
has been variationally solved in refs. \cite{ReiFeu04,FeuRei04} with the following ansatz for the
vacuum wave functional 
\bee
\label{Psi}
\Psi_\lambda [A] =\cN\cJ^{-\lambda}[A]\exp\left[-\frac{1}{2}\int d^3[xy]A_i^a(x)\omega(x,y)A_i^a(y)\right]
  \hk ,
\eeq
where  $\omega (x, y)$ is a variational kernel and $\lambda$ is a real
parameter. The choice $\lambda > 0$ seems to be appropriate since it
enhances field configurations near the Gribov horizon where $J [A] = 0$. It
turned out that in 1-loop approximation to the Dyson-Schwinger equations
arising from the minimisation of the energy density, the resulting 
infrared behaviour is independent of the value of $\lambda$. Instead, it is the occurrence of the Faddeev-Popov determinant in the Hamiltonian that is crucial to the Dyson-Schwinger equation \cite{ReiFeu04}. In order to investigate general features common to both the Coulomb and the Landau gauge, we now introduce a generating functional that facilitates the evaluation of expectation values of field operators in the Coulomb gauge\footnote{The integration of the path integral should be restricted to the fundamental modular region (FMR) $\Lambda\subset\Omega$ which is free of gauge copies. However, integrating over $\Omega$ will yield the same expectation values \cite{Zwa04}. Integration over $\Omega$ is understood in the following.}:
\bee
\label{ZCoulomb}
Z^{(C)}_\lambda [j] & = & \langle {\Psi} | \exp \left[ { \int d^3 x j^a_i ({ x}) A^a_i 
({ x})} \right] | {\Psi} \rangle
\nonumber\\
& = & {\cal N}^2 \int \mathcal{D} A \cJ^{1-2\lambda}[A]\exp \left[ - \int d^3 x \:\cL({ x}) +  \int
d^3 x j^a_i ({ x}) A^a_i ({ x}) \right] \; ,\hk 
\eeq
where
\bee
\label{LCoulomb}
 \cL({ x})= \int
d^3 x' 
 A^a_i ({ x} )\: \omega 
({ x}, { x}') A^a_i ({ x}')\; .
\eeq
Setting $\lambda=0$ w.l.o.g., as mentioned above, a reinterpretation of the Faddeev-Popov determinant by means of Grassmann valued ghost fields becomes feasible and one can subsequently use common Dyson-Schwinger techniques to derive the equations for the Green functions. 

From Eqs.\ (\ref{ZCoulomb}) and (\ref{ZLandau}) it is evident that apart from 
approximations, the expectation values in Landau and Coulomb gauge follow from the same generating functionals by merely swapping the dimension
(either $d=3$ or $d=4$) and the respective actions. The infrared behaviour in
the Coulomb gauge, in particular, is well described by a stochastic type of vacuum, as argued in \cite{ReiFeu04,Zwa03a}. That is, setting $\Psi[A]=1$ will yield the
correct infrared behaviour since it is dominated by the Faddeev-Popov
determinant, i.e.\ by the curvature in orbit space. This circumstance, called ``ghost dominance'', corresponds to setting $\lambda=0$ and $\cL=0$ in Eq.\ ({\ref{ZCoulomb}}). In the Landau gauge, ghost dominance was found as well \cite{LerSme02,AlkFisLla04}, i.e.\ setting $\cL_{YM}=0$ will not affect the solution in the infrared. One is led to the conclusion that the infrared behaviour of the solutions of Dyson-Schwinger equations are the same in Coulomb and Landau gauge, if we consider $d=3$ and $d=4$, respectively. Therefore, calculating moments of the following
generating functional that solely involves the Faddeev-Popov determinant,\footnote{Integrating in a compact region such as $\Omega$ ensures convergence of the path integral \cite{Zwa03a}.}

\bee
\label{Zboth}
Z^{(ir)}[j,\sigma,\bar{\sigma}] = {\cal N}^2 \int \cD A \int \cD [c\bar{c}]
 \exp \left[ \int d^d x \left( -\bar{c}\: D[A]\cdot\partial\: c + j \cdot A + \bar{\sigma} \cdot c + \bar{c} \cdot \sigma \right) \right]\; ,
\eeq

one recovers the correct expectation values of spatial components of field operators in the Coulomb gauge by setting $d=3$ and also the correct expectation values of field operators in the Landau gauge by choosing $d=4$, as far as the infrared is concerned.


\section{Infrared Analysis of Green functions}
\label{sec:analyt}

The truncated set of Dyson--Schwinger equations (DSEs) that govern the Green functions of the theory defined by Eq.\ (\ref{Zboth}) can be solved analytically in the infrared limit. To do so, we make the ansatz that the propagators obey power laws in the infrared, and determine the values of the exponents. The investigation of the vertex functions, in particular the ghost-gluon and the three-gluon vertex then follows by investigating the corresponding Dyson--Schwinger equations.

In the following it will be sensible to study Dyson--Schwinger equations in $d$-dimensional Euclidean spacetime. One may then specify to the Green functions of the Yang--Mills vacuum in the Landau gauge by setting $d=4$, or the Landau gauge high-temperature phase for $d=3$ \cite{Maas}. In the Coulomb gauge, one can derive Dyson--Schwinger equations for equal-time Green functions which corresponds to the choice $d=3$.

\subsection{Propagators}
\label{sec:props}
In refs.\ \cite{Zwa02,LerSme02}, a solution for the infrared behaviour of the propagators was previously obtained. We briefly review here the derivation of these results and critically analyse the approximations made. On the basis of our analysis we will have to discard one of the solutions found in ref.\ \cite{Zwa02}.

The crucial dynamical properties of Yang--Mills theory are accessible via the computation of its two-point Green functions, i.e.\ the ghost propagator,
\bee
\label{ghostdef}
D_G^{ab}(p):= \int d^dx\lla c^a(x)\bar{c}^b(y)\rra \e^{-ip\cdot (x-y)}=\delta^{ab}\frac{1}{g}\frac{G(p)}{p^2} \: ,
\eeq
as well as the gluon propagator,
\bee
\label{gluondef}
D^{ab}_{\mu\nu}(p):= \int d^dx\lla A_\mu^a(x)A_\nu^b(y)\rra \e^{-ip\cdot (x-y)}=\delta^{ab}t_{\mu\nu}(p)D_Z(p)=\delta^{ab}t_{\mu\nu}(p)\frac{Z(p)}{p^2} \: .
\eeq

\begin{figure}[t]
\ing[clip=]{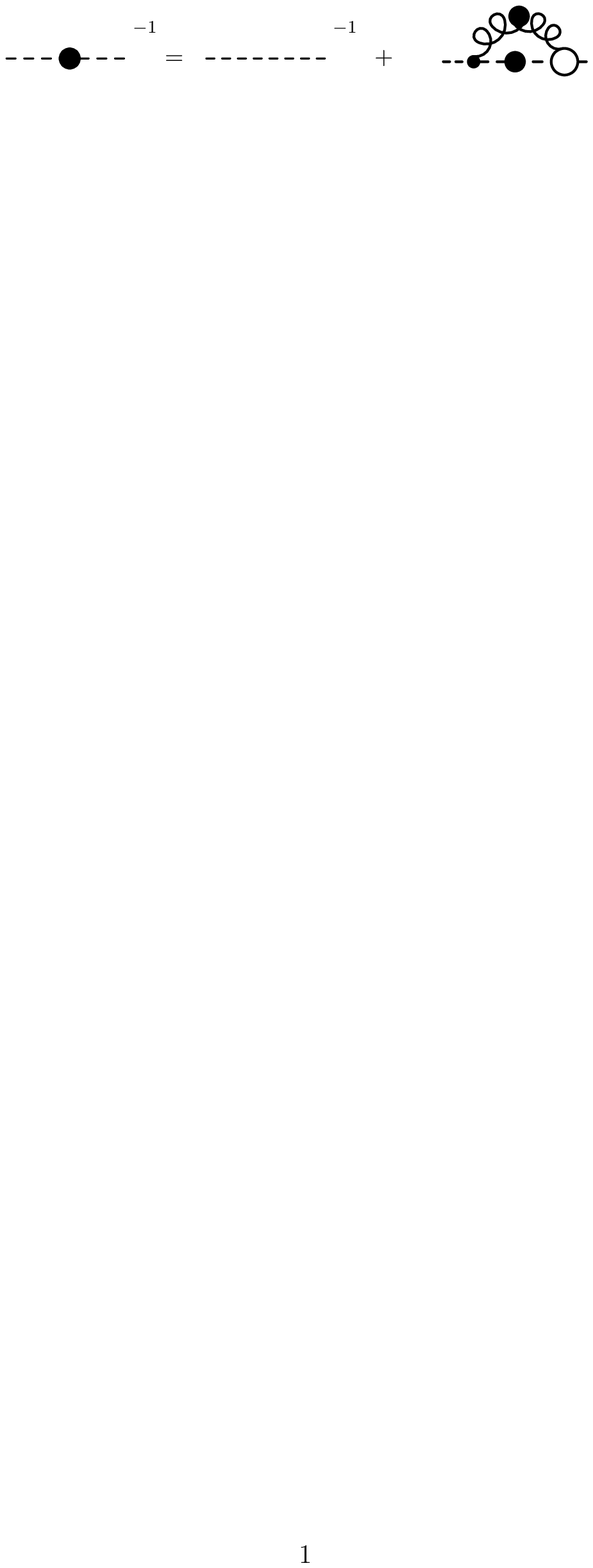}
\caption{The (complete) Dyson--Schwinger equation for the ghost propagator, denoted by a dashed line with a full blob. The curly line represents a connected gluon propagator and the vertex with an empty blob is a proper ghost-gluon vertex.}\label{fig:ghostDSE}
\end{figure}

These expectation values as well as the set of DSEs that entangles them with higher $n$-point Green functions can be derived from the generating functional, see e.g.\ \cite{FeuRei04,AlkSme00}. Part of the information about the infrared behaviour of the ghost and the
gluon propagator can be extracted from the ghost DSE. Using the bare
ghost-gluon vertex\footnote{A discussion of a more general form of the vertex
  will follow further below.}, it reads
\bee
G^{-1}(p)=\frac{1}{g}-gN_c\int  d^d\ellbar \: \left(1-(\hat{\ell}\cdot \hat{p})^2\right)D_Z(\ell)D_G(\ell-p)\: ,
\eeq
where the integral is represented by the self-energy diagram shown in Fig.\ \ref{fig:ghostDSE}. This integral bears an ultraviolet divergence for both three and four dimensions, due to the ultraviolet behaviour of the propagators. It can be conveniently subtracted by making use of the horizon condition \cite{Zwa91}, 
\bee
\label{horizon}
\lim_{p\rightarrow 0}G^{-1}(p)=0\: ,
\eeq
to find the finite expression
\bee
\label{renoghost}
G^{-1}(p)=G^{-1}(p)-G^{-1}(0)&=&gN_c\int  d^d\ellbar \: Z(\ell)(1-(\hat{\ell}\cdot \hat{p})^2)(D_G(\ell)-D_G(\ell-p))\: .
\eeq 
Aiming at the behaviour of $G(p)$ for $p\rightarrow 0$, it is instructive to assume that below some intermediate momentum scale $\xi$, the propagator dressing functions obey
\bee
\label{IIA}
G(p)\rightarrow G^{(ir)}(p)=\frac{B}{(p^2)^{\alpha_G}}\; ,\quad Z(p)\rightarrow Z^{(ir)}(p)=\frac{A}{(p^2)^{\alpha_Z}}\; , \quad p<\xi\; .
\eeq
As done in refs.\ \cite{Zwa02,LerSme02}, the integral in Eq.\ (\ref{renoghost}) can be analytically evaluated if we naively replace the propagators in the integrand by the power laws given by Eq.\ (\ref{IIA}), even though the integration is over all space. This procedure is referred to as  ``infrared integral approximation'' from now on. It leads to the following representation of the integral equation (\ref{renoghost}) in the infrared
\bee
\label{ghostDSE}
(p^2)^{\alpha_G}=(p^2)^{\frac{d-4}{2}-\alpha_G-\alpha_Z}AB^2N_cI_G(\alpha_G,\alpha_Z)+\Phi_G(p^2)\; ,
\eeq
where $I_G$ is a dimensionless number calculated in appendix A, see Eq.\ (\ref{IG}), and $\Phi_G$ is the error to account for the infrared integral approximation. Neglecting this error, one obtains the ``sum rule''
\bee
\label{sumrule}
\alpha_Z+2\alpha_G=\frac{d-4}{2}\; ,
\eeq
along with
\bee
\label{AB1}
AB^2N_cI_G(\kappa)=1 \; .
\eeq 

The infrared ghost exponent $\kappa:=\alpha_G$ can then be found by plugging
Eq.\ (\ref{AB1}) into the gluon DSE, as will be done below. 

\begin{figure}[t]
\centering
\ing[scale=0.77,clip=]{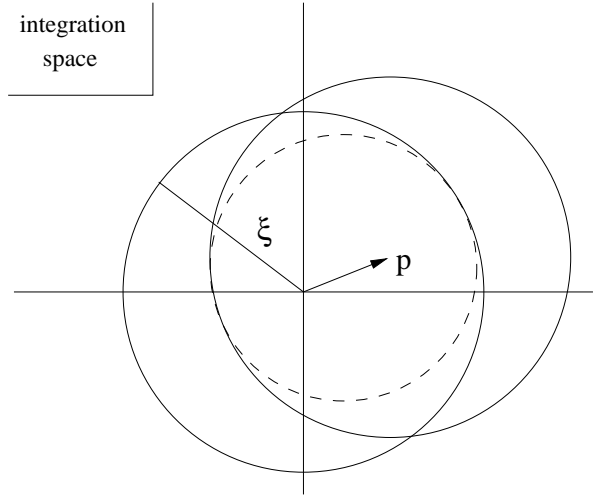}
\caption{Sketch of the integration space of (\ref{renoghost}) for $p<\xi$. Inside the full circles, the propagators are given by their infrared behaviour. Adding and subtracting the infrared power laws outside the circles makes it possible to evaluate the integrals analytically.}
\label{fig:infra}
\end{figure}

Before, however, let us make an
estimate for $\Phi_G$ in order to understand which values of $\kappa$ are at
all consistent with the sum rule (\ref{sumrule}) and when it is reasonable to
make use of the infrared integral approximation.
What happens when employing the replacements (\ref{IIA}) can be visualised by Fig.\ \ref{fig:infra}. For any generic self-energy type integral, 
\bee
\label{2ptgeneric}
I=\int d^d\ellbar\: D(\ell)D(\ell-p)\; ,
\eeq
the integration space can be divided into regions where the factors in the
integrand yield the infrared power law and ones where they do
not.\footnote{The following arguments can be applied as well to non-perturbative one-loop integrals of higher $n$-point functions.}  Here,
inside of a $d$-dimensional sphere of radius $\xi$ around the origin, one has
$D(\ell)=D^{(ir)}(\ell)$, and inside a sphere displaced by $p$, one finds
$D(\ell-p)=D^{(ir)}(\ell-p)$. The crucial point is that it is only inside of
these spheres, where the integrand may become non-analytic, i.e.\ the ghost
propagator has a pole at vanishing momentum, due to the horizon condition. As
long as $p\ll\xi$, we can always find a {\it third} sphere in the intersection of
the two others, see Fig.\ \ref{fig:infra}, inside of which all possible poles of the
integrand lie. If we then add and subtract $D^{(ir)}(\ell)D^{(ir)}(\ell-p)$ outside the third sphere, it becomes possible to represent
the original integral $I$ by a sum of integrals. Asymptotically, as $p\rarr 0$ and all three spheres intersect, this sum reads
\bee
I&\stackrel{{p\rarr 0}}{\longrightarrow}&\int_{\ell<\xi} d^d\ellbar\: D^{(ir)}(\ell)D^{(ir)}(\ell-p) +\int_{\ell\geq\xi} d^d\ellbar \:D^{(uv)}(\ell)D^{(uv)}(\ell-p)\nn\\&& +\int_{\ell\geq\xi} d^d\ellbar\: D^{(ir)}(\ell)D^{(ir)}(\ell-p)-\int_{\ell\geq\xi} d^d\ellbar \:D^{(ir)}(\ell)D^{(ir)}(\ell-p)\nn
\eeq
and we can define the first term plus the third term to be $I_0=\int d^d\ellbar\: D^{(ir)}(\ell)D^{(ir)}(\ell-p)=I-\Phi$, the infrared integral approximation of $I$. $I_0$ is an integral which regards the integrand factors as the infrared power laws (\ref{IIA}) over all space and ``captures'' any singularities of the original integrand. The error of the approximation, $\Phi=I-I_0$, integrates analytic functions and we can therefore expand it into a power series,
\bee\
\label{Phi}
\Phi(p^2)=\sum_{n=0}^\infty a_n(p^2)^n \; , \quad p\rarr 0\; ,
\eeq
which is finite at $p=0$. The integral $I_0$, on the other hand, diverges as $p\rarr 0$ for those values of the infrared exponents where $I_0$ exists. This can be understood from the convergence criteria for two-point integrals given in appendix A. $\Phi$ is then negligible for $p\rarr 0$ and one can set $I=I_0$. On the contrary, if
$p\ll\xi$ is not satisfied, the term $\Phi$ does have a substantial contribution, as can be seen, e.g., in numerical calculations in section \ref{sec:3gl}.

Returning to the ghost DSE (\ref{renoghost}), we note that the renormalisation plays an important role for the error $\Phi_G$ of the infrared integral approximation. Within the subtraction of the UV divergence the term with $n=0$ cancels in a power series such as (\ref{Phi}). We therefore find $\Phi_G(p^2)=\cO(p^2)$ and infer that we can neglect this term for $p\rarr 0$ in the sum $\Phi_G(p^2)+(p^2)^\kappa$ of Eq.\ (\ref{ghostDSE}) as long as 
\bee
\label{kapparange}
0<\kappa=\alpha_G<1 \: .
\eeq
The lower bound is due to the horizon condition (\ref{horizon}). On condition of the above relation, the sum rule (\ref{sumrule}) is
satisfied. For values of $\kappa$ with $\kappa\geq 1$, the
power series $\Phi_G$ does have to be taken into account, and one arrives at a
sum rule different from (\ref{sumrule}). However, those values are discarded
since they do not allow for Fourier transformation of the propagators.

\begin{figure}[t]
\ing[clip=]{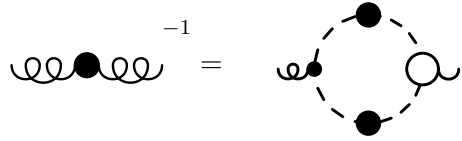}
\caption{The (truncated) Dyson--Schwinger equation for the gluon propagator. In the infrared, other terms are negligible.}
\label{fig:gluonDSE}
\end{figure}

Turning our attention to the gluon propagator DSE, we note that since the Faddeev-Popov determinant $\cJ[A]$ dominates the infrared, only the ghost loop has to be included. This has been found in the Landau gauge \cite{LerSme02,AlkFisLla04} as well as in the Coulomb gauge \cite{FeuRei04} for the equal-time gluon propagator. After contracting the gluon DSE with the transverse projector and taking the trace, we find \cite{Zwa02}
\bee
\label{gluonDSE}
Z^{-1}(p)=g^2N_c\frac{1}{(d-1)p^2}\int d^d\ellbar\:\ell^2\left(1-(\hat p\cdot\hat\ell)^2\right)D_G(\ell)D_G(\ell-p)\; ,
\eeq
see Fig.\ \ref{fig:gluonDSE}.
This integral is convergent in the ultraviolet. Employing the infrared integral approximation might introduce a spurious
ultraviolet divergence, depending on the value of $\kappa$. Of course, this
divergence has to cancel with the error $\Phi_Z$ of the approximation, since a choice of
the infrared behaviour of the integrand will not affect the
ultraviolet. Hence, there is no lower bound on $\kappa$ other than the horizon
condition. The upper bound given by Eq.\ (\ref{kapparange}) guarantees convergence of the integral
in the infrared. We then find
\bee
\label{IZdef}
(p^2)^{\alpha_Z}=(p^2)^{\frac{d-4}{2}-2\kappa}AB^2N_cI_Z(\kappa)+\Phi_Z(p^2)\: , \quad p\rarr 0\: ,
\eeq
where $I_Z(\kappa)$ is given in the appendix A, see Eq.\ (\ref{IZ}). The
error $\Phi_Z$ is completely negligible since it approaches a finite constant in the
infrared limit whereas the other terms diverge. This can be understood by noting $\alpha_Z=(d-4)/2-2\kappa<0$ for $\kappa>0$. Thus, for $p\rarr 0$, Eq.\ (\ref{IZdef})
reproduces the sum rule (\ref{sumrule}) and gives
\bee
\label{AB1IZ}
AB^2N_cI_Z(\kappa)=1\: .
\eeq
Along with Eq.\ (\ref{AB1}), this leads to
\bee
\label{detkappa}
I_G(\kappa)=I_Z(\kappa)\: ,
\eeq
the conditional equation for $\kappa$. For $d=4$, the Landau gauge case, only one unique solution lies in the
range given by Eq.\ (\ref{kapparange}), $\kappa^{(L)}\approx 0.595$
\cite{LerSme02,Zwa02}. The solution $\kappa^{(L)}=1$, claimed in \cite{Zwa02},
could not be confirmed.\footnote{The reason is that $\frac{1}{2}=\lim_{\kappa\rarr
    1}\left.\frac{I_Z(\kappa)}{I_G(\kappa)}\right|_{d=4}\neq\: \lim_{d\rarr 4}\left(\lim_{\kappa\rarr
    1}\frac{I_Z(\kappa)}{I_G(\kappa)}\right)=1$. Furthermore, for $\kappa=1$ the term $\Phi_G$ is not
negligible in Eq.\ (\ref{ghostDSE}).} In the case $d=3$, applicable for the
Coulomb gauge, we find two solutions, $\kappa^{(C)}_1\approx 0.398$ and
$\kappa^{(C)}_2=1/2$, in complete agreement with \cite{Zwa02}. Only one of
them, $\kappa^{(C)}_2$, is found using the angular approximation
\cite{FeuRei04}. Numerical calculations in \cite{FeuRei04} approximately
approach the other value $\kappa^{(C)}_1$. However, an improvement of the
numerical methods shows that $\kappa^{(C)}_2$ is also a stable solution \cite{EppReiSch06}. The
latter would result in a Coulomb potential that rises strictly linearly. Which
one of the solutions is energetically favoured is therefore an interesting
issue that is yet to be investigated.

\subsection{Ghost-gluon vertex}
The infrared behaviour of the ghost-gluon vertex in the Landau gauge of $SU(N_c)$ Yang-Mills theory for both $d=4$ and $d=3$ was investigated in Dyson-Schwinger studies \cite{dipl_Schleifenbaum,Sch+05} and in lattice calculations for $SU(2)$ and $d=4$ \cite{CucMenMih04, Ste+05}. It was found that its non-renormalisation which holds to all orders of perturbation theory \cite{Tay71}, remains valid in the non-perturbative regime. An appropriate question to ask is, to what extent does a finite dressing of the ghost-gluon vertex influence the numerical solution for the infrared exponent $\kappa$ of the propagators? 

Leaving aside the colour structure which is assumed to be that of the bare vertex, i.e.\ $f^{abc}$, we denote the proper reduced ghost-gluon vertex by $\Gamma_\mu(k;q,p)$ where $k$ is the outgoing gluon, $q$ the outgoing ghost and $p$ the incoming ghost momentum. Following ref.\ \cite{LerSme02}, a quite general ansatz for the ghost-gluon vertex is\footnote{Generally, there is another component along the gluon momentum which, however, has no contribution when contracted with a gluon propagator. Therefore, it can be discarded.} 
\bee
\label{GGZvertex}
\Gamma_\mu(k;q,p)=igq_\mu\sum_iC_i\left(\frac{k}{\sigma}\right)^{l_i}\left(\frac{q}{\sigma}\right)^{m_i}\left(\frac{p}{\sigma}\right)^{n_i}\: ,
\eeq
where the constraint $l_i+m_i+n_i=0, \forall i,$ guarantees the independence of the renormalisation scale $\sigma$, i.e.\ non-renormalisation of the vertex. It is readily shown \cite{LerSme02} that the sum rule (\ref{sumrule}) is not affected by a dressing of the ghost-gluon vertex such as (\ref{GGZvertex}), since it turns into 
\bee
\alpha_Z+2\alpha_G=\frac{d-4}{2}+\sum_i(l_i+m_i+n_i)\: .
\eeq
Further investigations in \cite{LerSme02} showed that the value for $\kappa$, determined by Eq.\ (\ref{detkappa}), only slightly depends on the values of $\{l_i,m_i,n_i\}$. 

Since neither the DSE studies \cite{dipl_Schleifenbaum,Sch+05} nor the lattice calculations \cite{CucMenMih04, Ste+05} show any infrared divergences, the dressing function of the ghost-gluon vertex must be some finite function. To investigate the consequences of a finite dressing function of the ghost-gluon vertex, let us assume, for simplicity, that it is given by a finite constant,
\bee
\label{constGGZ}
\Gamma_\mu(k;q,p)=C\Gamma^{(0)}_\mu(q)\; ,
\eeq
where $\Gamma^{(0)}_\mu(q)=igq_\mu$ is the bare ghost-gluon vertex. Then, the infrared analysis of the propagators can be performed in the same way as above. The ghost self-energy and the ghost loop are both multiplied by the constant $C$. In Eq.\ (\ref{detkappa}), this constant appears on both sides to one power and thus trivially cancels. Therefore, a constant dressing of the ghost-gluon vertex is completely irrelevant for the infrared behaviour of the propagators.

The question that arises is if a non-constant dressing of the ghost-gluon vertex might result in a change for the determining equation (\ref{detkappa}) of $\kappa$. The investigations in \cite{Sch+05} showed that after one iteration step of the ghost-gluon vertex DSE, the vertex remains approximately bare over the whole momentum range, i.e.\ $C\approx 1$. Also, the results in \cite{Sch+05} confirmed the well-known fact \cite{Tay71} that for vanishing incoming ghost momentum $p$, the ghost-vertex becomes bare in the Landau gauge.\footnote{This agrees with the corresponding Slavnov-Taylor identity in the Landau gauge.} It is essential for the proof given in ref.\ \cite{Tay71} that the gluon propagator is strictly transverse.  It has been argued that the same holds true for the Coulomb gauge \cite{FisZwa05}, where the gluon propagator is transverse as well. If we discard the irrelevant component of the vertex along the gluon momentum $k$, $\Gamma_\mu$ also becomes bare for vanishing outgoing ghost momentum $q$ \cite{LerSme02,dipl_Schleifenbaum}, i.e.\   
\bee
\label{symm}
\lim_{p\rarr 0}\Gamma_\mu(k;q,p)=\lim_{q\rarr 0}\Gamma_\mu(k;q,p)=\Gamma^{(0)}_\mu(q)\; .
\eeq

However, the infrared limits of the ghost and gluon momenta are generally not interchangeable.\footnote{In this context, one might note that the infrared limit of any tensor integral is non-trivial. Given an integral 
\bee
\label{tensint}
I_{\mu_1\mu_2\dots\mu_M}(\{p^{(i)}\})=\int d^d\ellbar\:\ell_{\mu_1}\ell_{\mu_2}\dots\ell_{\mu_M}\:f(\ell,\{p^{(i)}\})
\nn
\eeq
we can construct a tensor basis from the external scales $\{p^{(i)}\}$ and Lorentz invariant tensors. According to the Passarino-Veltman formalism, the above integral can then be expanded in this basis, which is nothing but solving a set of linear equations for the expansion coefficients in this basis. If one sets up a tensor expansion for finite $\{p^{(i)}\}$ and then tries to perform the infrared limit of a single external momentum, say $p^{(k)}\rarr 0$, the coefficient matrix becomes singular, and the tensor expansion is not well-defined. Instead, one can set $p^{(k)}=0$ from the beginning (if the integral exists here) and construct the tensor basis spanning a vector space which is of a lower dimension than originally. The expansion coefficients are then well-defined.} In particular, zero gluon momentum yields a dressing that is different from one, although quite close to it, as we will see. The following relation shall redefine $C$,
\bee
\label{Cdef}
C\:\Gamma^{(0)}_\mu(q)\equiv\lim_{p\rarr 0}\left(\lim_{k\rarr 0}\Gamma_\mu(k;q,p)\right)\neq\lim_{k\rarr 0}\left(\lim_{p\rarr 0}\Gamma_\mu(k;q,p)\right)=\Gamma^{(0)}_\mu(q) \: , \quad q\rarr 0\: .
\eeq

Does $C\neq 1$ or a non-constant $C$ affect Eq.\ (\ref{detkappa})? To see this, it is not necessary to get involved in a numerical calculation but we can argue qualitatively instead. Consider any loop integral that involves the ghost-gluon vertex.  Wherever it may appear in the loop diagram, the ghost-gluon vertex is always attached to ghost propagators. The integrand will be strongly enhanced for those loop momenta where the ghost propagator diverges, i.e.\ for $p\rightarrow 0$. Since the gluon propagator, on the other hand, is finite for all momenta, any infrared singularities in the integrand of the loop integral can actually be due to the ghost propagator only. The ghost-gluon vertex does not introduce any additional singularities, since its dressing function is finite. For the value of the integral, the singularities in the integrand will give the dominant contribution.  The only value of the dressing function of the ghost-gluon vertex that is relevant to the integral, is then the one where any of the ghost momenta vanish. According to Eq.\ (\ref{symm}), the vertex is bare in these limits. We can therefore infer that in any loop integral the bare ghost-gluon vertex will yield the correct result. This circumstance can thus be traced back to the horizon condition and the transversality of the gluon propagator.

\begin{figure}[t]
\ing[clip=]{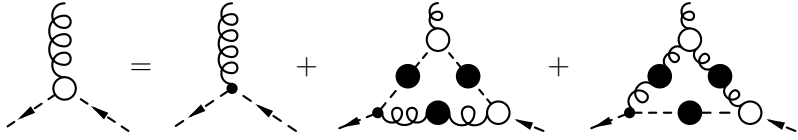}
\caption{The (truncated) Dyson--Schwinger equation for the ghost-gluon vertex.}
\label{fig:GGZDSE}
\end{figure}

Nevertheless, the constant $C$, defined by Eq.\ (\ref{Cdef}) is not entirely meaningless since the introduction of a running coupling, see section \ref{fixpt} below, makes use of it. One can actually analytically calculate $C$ by means of the DSE for the ghost-gluon vertex \cite{Sch+05}, see Fig.\ \ref{fig:GGZDSE},

\bee
\label{GGZDSE}
\Gamma_\mu(k;q,p)=\Gamma^{(0)}_\mu(q)+\Gamma^{(GGZ)}_\mu(k;q,p)+\Gamma^{(GZZ)}_\mu(k;q,p)\; .
\eeq

Here, $\Gamma^{(GGZ)}_\mu$ is a graph with two full ghost and one full gluon propagator in the loop,
\bee
\label{GGZ1loopGGZ}
\Gamma^{(GGZ)}_\mu(k;q,p)&=&-\frac{N_c}{2}\int d^d\ellbar\: 
\Gamma_\alpha^{(0)}(-\ell)D_{\alpha\beta}(\ell-q)\Gamma_\beta(q-\ell;-p,-\ell-k)\nn\\&&\hspace{2cm}D_G(\ell+k)\Gamma_\mu(k;\ell,\ell+k)D_G(\ell)
\: ,
\eeq
and $\Gamma^{(GZZ)}_\mu$ has two gluon and one ghost propagator in the loop, but involves a proper reduced three-gluon vertex $\Gamma_{\mu\nu\rho}$,
\bee
\label{GGZ1loopGZZ}
\Gamma^{(GZZ)}_\mu(k;q,p)&=&-\frac{N_c}{2}\int d^d\ellbar\: 
\Gamma_\alpha^{(0)}(q)D_G(\ell-q)\Gamma_\beta(\ell+k;q-\ell,p)
\nn\\&&\hspace{2cm}D_{\beta\rho}(\ell+k)\Gamma_{\mu\nu\rho}(k;\ell,-\ell-k)D_{\nu\alpha}(\ell)
\: .
\eeq

Since $\Gamma_\mu(k;q,p)$ exists in the limit $k\rarr 0$ \cite{Sch+05,CucMenMih04,Ste+05}, we set $k=0$ in the integrands which greatly simplifies the tensor structure of Eqs.\ (\ref{GGZ1loopGGZ}) and (\ref{GGZ1loopGZZ}). Furthermore, the proper ghost-gluon vertices that appear in the loop integrals are rendered bare, as discussed above. We then get
\bee
\label{GGZDSE1stk0}
\Gamma^{(GGZ)}_\mu(0;q,q)&=&ig^3C\frac{N_c}{2}\int d^d\ellbar\: 
\ell_\alpha D_{\alpha\beta}(\ell-q)q_\beta\ell_\mu D_G^2(\ell)
\: ,
\eeq
and 
\bee
\label{GGZDSE2ndk0}
\Gamma^{(GZZ)}_\mu(0;q,q)&=&g^2\frac{N_c}{2}\int d^d\ellbar\: 
q_\alpha q_\beta D_{\alpha\nu}(\ell)D_{\beta\rho}(\ell)D_G(\ell-q)\Gamma_{\mu\nu\rho}(0;\ell,-\ell)
\: .
\eeq
Naively, we would expect from ghost dominance in the infrared that the contribution (\ref{GGZDSE2ndk0}) is subdominant since it incorporates only one and not two ghost propagators, like (\ref{GGZDSE1stk0}). Using a bare three-gluon vertex, we can calculate both integrals for $q\rarr 0$ in the infrared integral approximation and indeed find that (\ref{GGZDSE2ndk0}) becomes negligible. The dominant part of the two is then (\ref{GGZDSE1stk0}), and it gives (see appendix A)
\bee
\label{I1def}
\lim_{q\rarr 0}\Gamma^{(GGZ)}_\mu(0;q,q)&=&\Gamma_\mu^{(0)}(q)\:g^2C\frac{N_c}{2}\int d^d\ellbar\: \ell_{\alpha\beta}(q)t_{\alpha\beta}(\ell-q)D_G^2(\ell)D_Z(\ell-q)\\
\label{constC}
&=&\Gamma_\mu^{(0)}(q)\: C\frac{\left( d-1 \right)}{d\left( 1 + 2\,\kappa  \right)}\frac{ \,\Gamma(d - 2\,\kappa )\,{\Gamma(1 + \kappa )}^2}
  { \,\Gamma(\frac{d}{2})\,{\Gamma(\frac{d}{2} - \kappa )}^2\,\Gamma(1 - \frac{d}{2} + 2\,\kappa )}
\eeq
which agrees exactly with the results of numerical calculations of (\ref{GGZDSE}) in this limit \cite{dipl_Schleifenbaum}
. Because of this agreement, we infer that the error introduced by the infrared integral approximation, employed for the calculation of (\ref{GGZDSE1stk0}) and (\ref{GGZDSE2ndk0}), vanishes.

If the dressed three-gluon vertex is included, see below, the graph (\ref{GGZDSE2ndk0}) has a substantial contribution to this limit of the ghost-gluon vertex. The calculation is then somewhat more involved, see appendix A, but one can extract the values for C in all cases at hand:
\bee
\label{Cresult}
C=\left\{
\begin{aligned}
1.108 & \quad \textrm{for}\,\,d=4\: ,\;\kappa=\kappa^{(L)}\approx 0.595 \\
1.089 & \quad \textrm{for}\,\,d=3\: ,\;\kappa=\kappa^{(C)}_1 \approx 0.398 \quad . \\
1   & \quad \textrm{for}\,\,d=3\: ,\;\kappa=\kappa^{(C)}_2 = \frac{1}{2}
\end{aligned} 
\right.
\eeq

It is quite remarkable that in the Coulomb gauge with the solution $\kappa^{(C)}_2=1/2$, the two non-trivial graphs that appear in the DSE for the ghost-gluon vertex show an exact mutual cancellation in the infrared gluon limit,
\bee
\label{cancel}
\lim_{q\rarr 0}\left(\Gamma^{(GGZ)}_\mu(0;q,q)+\Gamma^{(GZZ)}_\mu(0;q,q)\right)=0\: .
\eeq
Therefore, the interchangeability of limits is recovered in this case only and
the ghost-gluon vertex becomes bare in all infrared limits. Note also that the
results (\ref{Cresult}) are independent of $N_c$. The colour trace that occurs in the loop diagrams of Eq.\ (\ref{GGZDSE}) yields a factor of $N_c/2$, see Eqs. (\ref{GGZ1loopGGZ}) and (\ref{GGZ1loopGZZ}), but it cancels with the propagator coefficient term $AB^2=1/(I_GN_c)$.

\subsection{Three-gluon vertex}
\label{sec:3gl}
As mentioned in section \ref{formalism}, it is only the Faddeev-Popov
determinant that influences the infrared behaviour of Yang-Mills theory. The
solution obtained for the infrared exponents of the propagators was found to
be independent of the three-gluon vertex, in particular, since it does not
contribute to the ghost self-energy term or the ghost loop contribution to the gluon self-energy. In the Coulomb
gauge, we have seen that any of the vacua  $\Psi_\lambda[A]$ given by Eq.\
(\ref{Psi}) minimises the energy w.r.t.\ $\lambda$, evaluated to one-loop order in the DSE (two-loop order in the diagrams for the energy). The question is how the three-gluon vertex changes in the infrared for different values of $\lambda$ without resorting to the one-loop approximation.

The full three-gluon vertex is defined as 
\bee
\Gamma^{abc}_{\mu\nu\rho}(x,y,z)=\lla\Psi|A^a_\mu(x)A^b_\nu(y)A^c_\rho(z) |\Psi\rra \; .
\eeq
In the particular case of $\lambda=1/2$ it is found that
\bee
\label{vertexzero}
\int\cD A A_\mu A_\nu A_\rho \e^{-\int A\omega A}=0
\eeq
by symmetry. Hence, the three-gluon vertex vanishes for $\psi_{1/2}$ \cite{FeuRei04}. Now consider the case $\lambda \neq 1/2$. In \cite{ReiFeu04} it has been found that the Faddeev-Popov determinant can be written to one-loop order as

\begin{equation}
\mathcal{J}[A]= \exp \left[-\int d^d[xx'] \chi^{ab}_{ij}(x,x') A_i^a(x)A_j^b(x')\right]
\end{equation}

with $\chi$ being the curvature of the space of gauge orbits, i.e.\ the ghost
loop contribution to the gluon self-energy. In this form, 
$\cJ$ merely modifies the inverse gluon propagator $\omega$ that occurs in Eq.\ (\ref{ZCoulomb}) according to

\begin{equation}
\omega(x,x')\longrightarrow\Omega(x,x') := \omega(x,x') - (2\lambda -1)\chi(x,x') \; .
\end{equation}

The subsequent determination of $\Omega$ by minimising the vacuum energy guarantees that $\Omega$ always is the same function, regardless of the $\lambda$ chosen. In particular, $\Omega=\omega$ for $\lambda = 1/2$. Therefore, all expectation values are the same to one-loop order in the equation of motion, so the three-gluon vertex vanishes for any $\lambda$.

On the other hand, without the use of the one-loop approximation, $\lambda\neq 1/2$ will give a non-zero three-gluon vertex, in contrast to Eq.\ (\ref{vertexzero}), as will be shown. Thus, the three-gluon vertex shows great sensitivity to the choice of the vacuum wave functional $\psi_\lambda$, a behaviour not exhibited to one-loop order. Making the choice $\lambda=0$ permits the standard representation of the Faddeev-Popov determinant by ghosts. In the following the infrared behaviour of the three-gluon vertex for this case will be explored following the treatment in \cite{dipl_Leder}.

\begin{figure}[t]
\centering
\includegraphics[width=10cm,clip=]{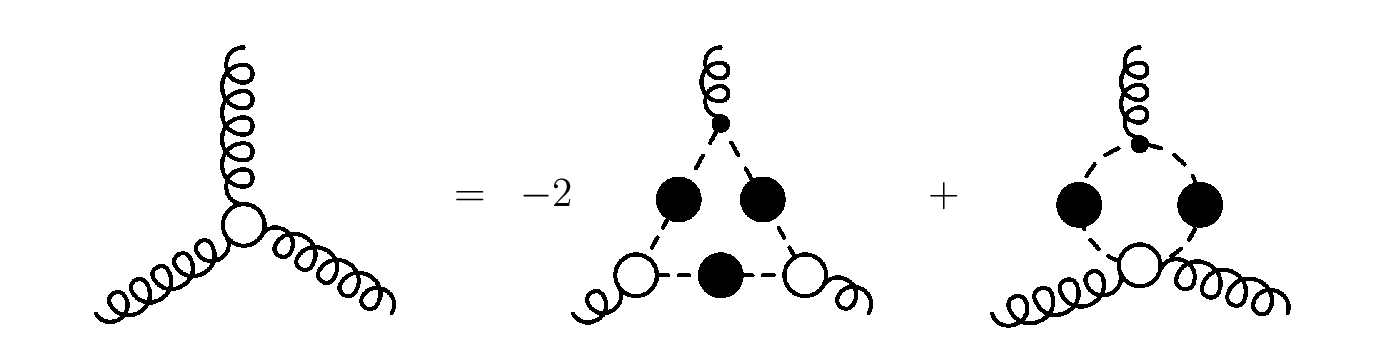}
\caption{The (complete) DSE for the three-gluon vertex derived from the
  generating functional given in Eq.\ (\ref{ZCoulomb}).}
\label{DSE diag}
\end{figure}

The Dyson-Schwinger equation for the three-gluon vertex is derived in Appendix
B and depicted diagrammatically in Fig. \ref{DSE diag}. Its complete form
comprises a diagram with the unknown two-ghost-two-gluon vertex which is
truncated here. The finite ghost-gluon vertex appears here in a loop integral and can
therefore be set to its bare value throughout, according to the discussion
given below Eq.\ (\ref{Cdef}) in the last subsection. Assuming tree-level colour
structure for all of the correlation functions and Fourier transforming the
truncated DSE, one arrives at
\bee
\label{DSEmomentum}
\hspace{-0.6cm}\Gamma_{\mu\nu\rho}(p_1,p_2,p_3) = N_c \int d^d\ellbar\; D_G(k) D_G(p_3+k) D_G(k-p_2)
 \Gamma^{(0)}_\mu(\ell) \Gamma^{(0)}_\nu(\ell-p_2)\Gamma_\rho^{(0)}(\ell+p_1)
\eeq
where the outgoing momenta obey the conservation law 
\begin{equation}
p_1 + p_2 + p_3 = 0 \; .
\end{equation}

The vertex given by Eq.\ (\ref{DSEmomentum}) is projected onto the tensor subspace spanned by the tensor components of the tree-level vertex. Due to Bose symmetry, the coefficient functions of these six components are all the same, but their signs alternate as the vertex without the colour structure is antisymmetric under gluon exchange. One finds

\begin{equation}\label{tensor}
\begin{aligned}
\Gamma_{\mu\nu\rho}(p_1,p_2,p_3) = & - i(p_2)_\mu \delta_{\nu\rho} F (p_2^2,p_1^2,p_3^2) && + i(p_3)_\mu \delta_{\nu\rho} \; 
F (p_3^2,p_1^2,p_2^2) \\
& + i(p_1)_\nu \delta_{\mu\rho} F (p_1^2,p_2^2,p_3^2) && - i(p_3)_\nu \delta_{\mu\rho} \; F (p_3^2,p_2^2,p_1^2) \\
& - i(p_1)_\rho \delta_{\mu\nu} F (p_1^2,p_3^2,p_2^2) && + i(p_2)_\rho \delta_{\mu\nu} \; F (p_2^2,p_3^2,p_1^2)\; . \\
\end{aligned}
\end{equation} 

Equating Eq.\ (\ref{DSEmomentum}) with (\ref{tensor}) and contracting with these six tensors, yields a set of six linear equations for $F$, the solution of which reads

\begin{multline}\label{formfactor}
F(p_1^2,p_2^2,p_3^2)= \frac{-N_c}{10(p_1^2 p_2^2 - (p_1\cdot p_2)^2)} \int
d^d\ellbar \: D_G(\ell) D_G(p_3+\ell)D_G(\ell-p_2) \\
\left( (p_1^2 + p_1\cdot p_2)(-2J_2-J_4+3J_5) +(p_2^2 + p_1\cdot p_2)(2J_1-3J_3+J_6)\right)
\end{multline}
 
where

\begin{equation}\label{J}
\begin{aligned}
J_1:=& (p_1\cdot \ell)\ell^2 & J_2:=& (p_2\cdot \ell)\ell^2 & J_3:=& (p_2\cdot \ell)(p_1\cdot \ell)\\
J_4:=& p_2^2\ell^2 & J_5:=& (p_2\cdot \ell)^2 & J_6:=& (p_1\cdot p_2)\ell^2 \; .
\end{aligned}
\end{equation}

\begin{figure}
\includegraphics[scale=0.4,clip=]{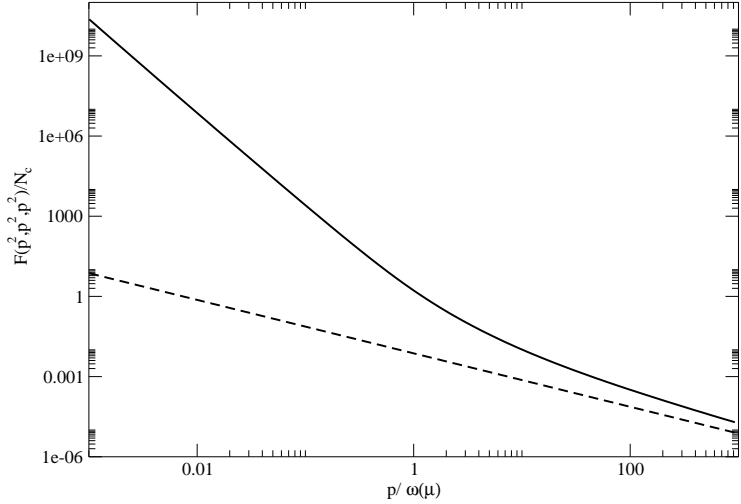}
\caption{Dressing function of the Coulomb gauge proper three-gluon vertex at the symmetric point. The dashed curve shows in contrast the perturbative case where the propagators in the loop are bare, i.e.\ $\kappa=0$.}
\label{fig:ZZZ}
\end{figure}

The integral (\ref{formfactor}) depends only on the ghost propagator in this
truncation, and despite the infrared enhancement of the latter it is
convergent. The numerical calculation of the form factor $F$ at the symmetric
point,  where $p_1^2=p_2^2=p_3^2=:p^2$, shows a strong infrared divergence,
see Fig.\ \ref{fig:ZZZ}. For the ghost propagator we used the numerical
results of ref.\  \cite{FeuRei04} where $\kappa=0.425\approx\kappa^{(C)}_1$. A
fit to the data in Fig.\  \ref{fig:ZZZ} yields an infrared power law such as $F(p^2)
\sim (p^2)^{-1.77}$. If we employ the infrared integral approximation to
calculate the symmetric point analytically, a power law behaviour can be
extracted as well.  By momentum scaling of the integration variable in Eq.\
(\ref{formfactor}), $\ell \rightarrow \lambda \ell$, one finds  $F(p^2)
\sim (p^2)^{d/2-2-3\kappa}$. Plugging in the value $\kappa=0.425$ obtained numerically in \cite{FeuRei04} gives an
infrared exponent of $-1.775$ which agrees (within errors) with the numerical
result. This shows that the infrared integral approximation becomes exact as the
external momenta vanish. Only in the ultraviolet there is a deviation from the infrared power law which can be clearly explained by the error of the approximation. At large momenta the vertex vanishes which complies with asymptotic freedom since in this approximation there is no tree-level vertex, due to Eq.\ (\ref{LCoulomb}).

The above results for the infrared behaviour of the Coulomb gauge three-gluon
vertex can be generalised to any value of $\kappa$ and any
dimension $d$. Therefore, we can also make statements about the Landau
gauge. Noting that $d/2-2-3\kappa=\alpha_Z-\alpha_G$, see Eq.\
(\ref{sumrule}), we find

\begin{equation}\label{3gl_powerlaw}
F(p^2) \sim \frac{1}{(p^2)^{\alpha_G-\alpha_Z}}
\end{equation}

With the analytical results for $\kappa$ in Coulomb ($d=3$) as well as in
Landau ($d=4$) gauge, this yields

\bee
\label{Fresult}
F(p^2)=\left\{
\begin{aligned}
\frac{1}{(p^2)^{1.775}} & \quad \textrm{for}\,\,d=3\: ,\;\kappa=\kappa^{(C)}_1\approx 0.398 \\
\frac{1}{(p^2)^{2}} & \quad \textrm{for}\,\,d=3\: ,\;\kappa=\kappa^{(C)}_2 = \frac{1}{2}\quad . \\
\frac{1}{(p^2)^{1.785}} & \quad \textrm{for}\,\,d=4\: ,\;\kappa=\kappa^{(L)} \approx 0.595 
\end{aligned}
\right.
\eeq

The Landau gauge result agrees exactly with ref.\ \cite{AlkFisLla04}. 

Another interesting kinematic point is the one where one of the gluon momenta, say $p_1$, is set to zero while the others remain finite. Trying to calculate this point from Eq.\ (\ref{tensor}) by setting $p_1=0$, the projections onto the tensor components fail as the determinant of the coefficient matrix that defines the tensor expansion vanishes in this case. It is advisory to impose $p_1=0$ in the DSE (\ref{DSEmomentum}), 
\bee
\label{ZZZDSEk0full}
\Gamma_{\mu\nu\rho}(0,p,-p)=-ig^3N_c\int d^d\ellbar\:\ell_\mu(\ell-p)_\nu\ell_\rho D_G^2(\ell)D_G(\ell-p)\: .
\eeq
One can then realize that this integral exists. It can be expanded into a
tensor basis constructed by the only scale $p$ and Lorentz invariant tensors,
i.e.\
$\{p_\mu\delta_{\nu\rho},p_\nu\delta_{\rho\mu},p_\rho\delta_{\mu\nu}\}$.
However, the only component that survives the transverse projections of the
gluon propagators attached to the legs of $\Gamma_{\mu\nu\rho}$ with finite momenta, is obviously $p_\mu\delta_{\nu\rho}$. Thus, we can write
\bee
\label{ZZZDSEk0}
\Gamma_{\mu\nu\rho}(0,p,-p)=-ig^3N_cp_\mu\delta_{\nu\rho}\frac{1}{(d-1)p^2}p_\alpha t_{\beta\gamma}(p)\int d^d\ellbar\:\ell_\alpha\ell_\beta\ell_\gamma D_G^2(\ell)D_G(\ell-p)+\dots
\eeq
where the ellipsis represents irrelevant tensor components which shall be discarded henceforth. Using the infrared integral approximation, we find a finite expression, see $I_3$ in appendix A, which makes the three-gluon vertex
\bee
\label{ZZZk0}
\Gamma_{\mu\nu\rho}(0,p,-p)=-iB^3N_cp_\mu\delta_{\nu\rho}\frac{I_3}{(p^2)^{\alpha_G-\alpha_Z}}\; , \quad p\rarr 0\: .
\eeq
The error can be ignored due to the infrared enhancement of (\ref{ZZZk0}).

In view of the strong infrared divergence of the three-gluon vertex, one has to check for ghost dominance in the propagator DSEs, which simply states that one is to count the infrared exponents of the propagators in the loop \cite{Zwa03a}. However, the vertex function have to be taken into account, too \cite{AlkFisLla04}. The infrared power law of the three-gluon vertex (\ref{3gl_powerlaw}), expresses that the vertex dressing replaces the infrared exponent of a gluon by that of a ghost propagator, for any dimension $d$. The infrared hierarchy of terms in the gluon DSE remains untouched, since even with the dressing of three-gluon vertex, terms involving it remain subleading in the infrared. E.g., the gluon loop, which has an infrared exponent of $d/2-2-2\alpha_Z$ with a bare three-gluon vertex, attains an infrared power law with the exponent $d/2-2-\alpha_Z-\alpha_G$ if the vertex is dressed. Clearly, this term is still subleading w.r.t.\ the ghost loop which bears an infrared exponent of $d/2-2-2\alpha_G$. 


\subsection{Infrared fixed point of the running coupling}
\label{fixpt}

A renormalisation group invariant that qualifies as a non-perturbative running
coupling can be extracted from the ghost-gluon vertex and is given by \cite{CucMenMih04}
\bee
\alpha(p^2)=\alpha_0\:\frac{\cZ_3(p^2)\widetilde{\cZ}_3^2(p^2)}{\widetilde{\cZ}_1^2(p^2)}\; .
\eeq
Here, $\alpha_0=g_0^2/4\pi$ is the bare coupling constant and $\cZ^{1/2}_3(p^2)$, $\widetilde{\cZ}^{1/2}_3(p^2)$ are the gluon and ghost field renormalisation functions, respectively. According to the definition of the renormalisation function $\widetilde{\cZ}_1(p^2)$ of the ghost-gluon vertex in \cite{CucMenMih04}, we find that $\widetilde{\cZ}^{-1}_1(0)=C$, with $C$ given by Eq.\ (\ref{Cresult}).\footnote{Note that we have used the definition of $C$ in Eq.\ (\ref{Cdef}). With the infrared limits not being interchangeable generally, an alternative renormalisation prescription of the ghost-gluon vertex would give $C=1$.} In the Landau gauge ($d=4$) this leads to an infrared fixed point $\alpha_c\equiv\alpha(0)$, 
\bee
\alpha_c^{(L)}=\left.\frac{AB^2C^2}{4\pi}\right|_{\kappa=\kappa^{(L)}}\approx\left.\frac{8.915}{N_c}C^2\right|_{\kappa=\kappa^{(L)}}\approx\frac{10.94}{N_c}\; .
\label{infrafixLandau}
\eeq
Up to the vertex correction $C^2$, this value was found in \cite{LerSme02}. Although the result (\ref{infrafixLandau}) holds for both Landau and the interpolating gauges \cite{FisZwa05}, the Coulomb gauge limit reveals an infrared fixed point different from it, even if $C=1$. By definition \cite{FisZwa05} one gets
\bee
\alpha_c^{(C)}=\left.\frac{4AB^2C^2}{3\pi}\right|_{\kappa=\kappa^{(C)}} \: .
\label{infrafixCoulomb}
\eeq
With the values for the integral $I_G(\kappa)$ at $d=3$ for the solution $\kappa^{(C)}_1=0.398$, and the vertex correction $C^2=1.187$, this yields
\bee
\left.\alpha_c^{(C)}\right|_{\kappa=\kappa^{(C)}_1}\approx\left.\frac{11.99}{N_c} C^2\right|_{\kappa=\kappa^{(C)}_1}\approx\frac{14.21}{N_c} \: .
\eeq
For the other Coulomb gauge solution, where $\kappa^{(C)}_2=1/2$ and the vertex correction vanishes, we find
\bee
\left.\alpha_c^{(C)}\right|_{\kappa=\kappa^{(C)}_2}=\frac{16\pi}{3N_c}\approx\frac{16.76}{N_c} \: .
\eeq
Among the interpolating gauges, the running coupling appears to have an infrared fixed point that changes discontinuously in the Coulomb gauge limit.

A second possibility for a definition of a running coupling is given by the instantaneous Coulomb potential which is singular in the infrared \cite{Zwa03a}. These two choices clearly disagree on the description of long-range interactions. It is known that the Coulomb string tension is an upper bound for the string tension extracted from the Wilson loop \cite{Zwa03b}.

\section{Summary and Conclusions}
\label{summary}
\bi

\no
We have studied the infrared limit of the ghost and gluon propagators as well as
of the ghost-gluon and three-gluon vertices to 1-loop order in both Coulomb and
Landau gauge assuming ghost (loop) dominance in the infrared. From the
Dyson--Schwinger equations we have found that there is a unique infrared
exponent in the Landau gauge, while there are two different exponents in the Coulomb
gauge corresponding presumably to different minima of the energy density. It
would be interesting to determine which one corresponds to the absolute minimum
of the energy density. In the Coulomb gauge for one of the infrared exponents we
found an exact cancellation between the two loop-diagrams of the DSE for the
ghost-gluon vertex. This vertex is infrared finite in both Landau and Coulomb
gauge. We have also shown, that a finite dressing of the ghost gluon vertex does
not modify the infrared exponents. The three-gluon vertex was found to be
infrared divergent with approximately the same infrared exponent in Coulomb and
Landau gauges. Furthermore, the infrared divergence of the three-gluon vertex
does not spoil the infrared dominance of the ghost loops over the gluon loops.
We also calculated the three-gluon vertex numerically in Coulomb gauge over the
whole momentum range for the symmetric point. The numerical result reproduces the infrared behaviour found
analytically. For one vanishing gluon momentum, we determined the three-gluon vertex for any dimension. Furthermore, the three-gluon vertex turned out to be quite sensitive to the specific form of the Yang-Mills vacuum wave functional, and
therefore a lattice calculation of this quantity would be of great interest.
Finally, we have also calculated the infrared fixed point of the running coupling
and found a larger value in Coulomb gauge than in Landau gauge.

\section*{Acknowledgements}
Part of this work was supported by DFG-Re856/6-1 and the Europ\"aisches Graduiertenkolleg. We are grateful to Claus Feuchter, Christian S.\ Fischer and Peter Watson for valuable discussions.

\appendix
\renewcommand{\theequation}{A\arabic{equation}}
\setcounter{equation}{0}  

\section*{Appendix A: Two-point integrals}
\label{2ptint}
Here we sketch the derivation of the integrals necessary to calculate the Feynman graphs in the infrared integral approximation. The two-point integrals,
\begin{equation}
 \Xi_m(\alpha,\beta):=\int\frac{d^{d}\ellbar\:(\ell\cdot q)^m}{(\ell^{2})^{\alpha}((\ell-q)^{2})^{\beta}}\quad , \alpha ,\beta\in\mathbb{R},m\in\mathbb{N}
 \: ,
\label{eq:2ptint}
\end{equation}
 which are encountered in the calculations can
be shown to be homogeneous functions of the momentum $q$. By a scaling of
the integration variable, $\ell\rightarrow\lambda\ell$, one readily finds that since the two-point integral can only depend on the scale, it should obey $\Xi_m(\alpha,\beta) \sim (q^2)^\kappa$ and the exponent of
the power law can be determined to be
\bee
\label{kappa}
\kappa=d/2-\alpha-\beta+m\:.
\eeq

After applying the usual trick of introducing Feynman parameters,

\begin{equation}
\frac{1}{C_{1}^{\alpha}C_{2}^{\beta}}=\int_{0}^{1}dxdy\delta(x+y-1)\:\frac{x^{\alpha-1}y^{\beta-1}}{(xC_{1}+yC_{2})^{\alpha+\beta}}\frac{1}{B(\alpha,\beta)}
\: ,
\end{equation}

where $B(\alpha,\beta)$ is the Euler beta function, we can shift the integration variable, $\ell\rarr\ell-yq$. The integrand then depends on $\ell^2$ only and we can integrate it out. For $m=0,1,2,3$ we need the following standard integrals:
\bee
\int\frac{\dbar^{d}\ell}{(\ell^{2}+\Delta)^{n}}&=&\frac{1}{(4\pi)^{d/2}}\frac{\Gamma(n-d/2)}{\Gamma(n)}\left(\frac{1}{\Delta}\right)^{n-d/2} \\
\int\frac{\dbar^{d}\ell\:\ell_\mu\ell_\nu}{(\ell^{2}+\Delta)^{n}}&=&\frac{1}{2}\:\delta_{\mu\nu}\frac{1}{(4\pi)^{d/2}}\frac{\Gamma(n-d/2)}{\Gamma(n)}\left(\frac{1}{\Delta}\right)^{n-d/2}
\: .
\eeq

Integrals with an odd number of vectors $\ell$ in the numerator vanish by symmetry. For our purposes, we have $\Delta=xyq^2$. The Feynman integrals can be straightforwardly computed using the identity
\bee
\int_0^1dxdy\delta(x+y-1)x^{\alpha-1}y^{\beta-1}=B(\alpha,\beta)
\eeq
for the beta function. One then finds the results

\bee
\Xi_0(\alpha,\beta)= \frac{1}{(4\pi)^{d/2}}\frac{\Gamma(d/2-\alpha)\Gamma(d/2-\beta)\Gamma(\alpha+\beta-d/2)}{\Gamma(\alpha)\Gamma(\beta)\Gamma(d-\alpha-\beta)}(q^{2})^{d/2-\alpha-\beta}\: ,
\label{eq:2pt}
\eeq
\bee
\Xi_1(\alpha,\beta) = \frac{1}{(4\pi)^{d/2}}\frac{\Gamma(d/2-\alpha+1)\Gamma(d/2-\beta)\Gamma(\alpha+\beta-d/2)}{\Gamma(\alpha)\Gamma(\beta)\Gamma(d-\alpha-\beta+1)}(q^{2})^{d/2-\alpha-\beta+1}\; ,
\label{eq:2ptNum1}
\eeq
\bee
\Xi_2(\alpha,\beta)\nonumber &=&
 \frac{1}{(4\pi)^{d/2}}\frac{\Gamma(d/2-\alpha+2)\Gamma(d/2-\beta)\Gamma(\alpha+\beta-d/2)}{\Gamma(\alpha)\Gamma(\beta)\Gamma(d-\alpha-\beta+2)}(q^{2})^{d/2-\alpha-\beta+2} \\
&& \hspace{-1cm}+ \frac{1}{2} \frac{1}{(4\pi)^{d/2}}\frac{\Gamma(d/2-\alpha+1)\Gamma(d/2-\beta+1)\Gamma(\alpha+\beta-d/2-1)}{\Gamma(\alpha)\Gamma(\beta)\Gamma(d-\alpha-\beta+2)}(q^{2})^{d/2-\alpha-\beta+2} \; ,
\label{eq:2ptNum2}
\eeq
\bee
\Xi_3(\alpha,\beta) \nonumber &=&
 \frac{1}{(4\pi)^{d/2}}\frac{\Gamma(d/2-\alpha+3)\Gamma(d/2-\beta)\Gamma(\alpha+\beta-d/2)}{\Gamma(\alpha)\Gamma(\beta)\Gamma(d-\alpha-\beta+3)}(q^{2})^{d/2-\alpha-\beta+3}\\ 
&& \hspace{-1.5cm}+ \frac{3}{2}\frac{1}{(4\pi)^{d/2}}\frac{\Gamma(d/2-\alpha+2)\Gamma(d/2-\beta+1)\Gamma(\alpha+\beta-d/2-1)}{\Gamma(\alpha)\Gamma(\beta)\Gamma(d-\alpha-\beta+2)}(q^{2})^{d/2-\alpha-\beta+3}\; .
\label{eq:2ptNum3}
\eeq

The above formulae are valid for those values of $\alpha$, $\beta$, $m$ only for which the integrals converge. At $\ell=q$, a pole is integrable as long as $\beta<d/2$. On the other hand, the infrared convergence at $\ell=0$ depends on $m$:
\bee
\label{IRconv}
\alpha < \left\{
\begin{array}{cc}
d/2+ m/2 & \quad \textrm{for even}\,\, m \\
d/2+m/2+1/2 & \quad \textrm{for odd}\,\, m 
\end{array}\; .
\right.
\eeq

One can relate this inequality to the requirement that the arguments of the first gamma functions in the numerators of (\ref{eq:2pt}), (\ref{eq:2ptNum1}), (\ref{eq:2ptNum2}) and (\ref{eq:2ptNum3}) be positive. The ultraviolet convergence is contained in the third gamma function of the numerators. For convergence, the relation
\bee
\label{UVconv}
\alpha+\beta > \left\{
\begin{array}{cc}
d/2+ m/2 & \quad \textrm{for even}\,\, m \\
d/2+m/2-1/2 & \quad \textrm{for odd}\,\, m 
\end{array}
\right.
\eeq
has to be satisfied. Obviously, odd values of $m$, compared to even values, work ``in favour'' of convergence both in the infrared and in the ultraviolet, due to the angular integration. It is worthwhile examining two-point integrals that do not converge. Making the usual replacement 
\bee
\ell\cdot q=\frac{1}{2}(\ell^2+q^2-(\ell-q)^2)
\eeq
in any of the convergent integrals $\Xi_m$ with $m>0$, one can express these in terms of a sum of $\Xi_0$ integrals, but generally encounters both IR and UV divergences. Curiously, if we use the regular result (\ref{eq:2pt}) for the divergent $\Xi_0$ integrals anyhow, the sum will yield the correct result given by the direct formulae (\ref{eq:2ptNum1}), (\ref{eq:2ptNum2}) or (\ref{eq:2ptNum3}), as can be checked. This indicates that divergent two-point integrals can be written as a regular part given by the formulae calculated here, plus the divergence which may cancel with another integral of the same kind.

This circumstance is of great use for the ghost DSE, where the subtraction of $G^{-1}(0)$ removes the UV divergence and we can calculate $I_G(\kappa)$ in Eq.\ (\ref{ghostDSE}) by
\bee
\label{IG}
I_G(\kappa)&=&(p^2)^{-\kappa}\left(\Xi_2(\frac{d}{2}-2\kappa,1+\kappa)/p^2-\Xi_0(\frac{d}{2}-2\kappa-1,1+\kappa)\right)_\textrm{reg.}\nn\\
&=&- \frac{4^\kappa (d-1)}{(4\pi)^{d/2+1/2}} \,\frac{\Gamma(\frac{d}{2} - \kappa )\,\Gamma(-\kappa )\,
      \Gamma(\frac{1}{2} + \kappa )}{\Gamma(\frac{d}{2} - 2\,\kappa )\,\Gamma(1 + \frac{d}{2} + \kappa )}
\eeq
The same result was found in \cite{Zwa02} where subtractions of divergences were circumvented.

The integral that occurs in the gluon DSE (\ref{gluonDSE}) is essentially $I_Z(\kappa)$ defined by Eq.\ (\ref{IZdef}), and can be calculated to give
\bee
\label{IZ}
I_Z(\kappa)&=&(p^2)^{-\kappa}\frac{1}{d-1}\left(\Xi_0(\kappa,1+\kappa)-\Xi_2(1+\kappa,1+\kappa)/p^2\right) \nn\\
&=&\frac{1}{2(4\pi)^{d/2}}\,\frac{{\Gamma(\frac{d}{2} - \kappa )}^2\,\Gamma(1 - \frac{d}{2} + 2\,\kappa )}
  {\Gamma(d - 2\,\kappa )\,{\Gamma(1 + \kappa )}^2}\: .
\eeq
Note that although both integrals in (\ref{IZ}) have an infrared divergence at $\ell=q$, the sum is regular for those values of $\kappa$ in (\ref{kapparange}), as can be seen by shifting $\ell\rarr\ell-q$. In the ultraviolet, (\ref{IZ}) only converges as long as $\kappa>d/4-1/2$. However, such a divergence will necessarily cancel with the error $\Phi_Z$, as discussed in section \ref{sec:props}. Regardless, the solutions for both the Coulomb and the Landau gauge yield $\kappa>d/4-1/2$.

For the calculation of the infrared limit of the ghost-gluon vertex, we encounter the integral in (\ref{constC}). With 
\bee\ell_{\alpha\beta}(q)t_{\alpha\beta}(\ell-q)=\frac{\ell^2}{(\ell-q)^2}-\frac{(\ell\cdot q)^2}{q^2(\ell-q)^2}\eeq
we find, using Eq.\ (\ref{AB1IZ}),
\bee
\lim_{q\rarr 0}\Gamma^{(GGZ)}_\mu(0;q,q)&=&\Gamma_\mu^{(0)}(q)\:Cg^2\frac{N_c}{2}AB^2\left(
\Xi_1(1+2\kappa,\frac{d}{2}-2\kappa)-\Xi_3(2+2\kappa,\frac{d}{2}-2\kappa)/q^2
\right)\nn\\
&=&\Gamma_\mu^{(0)}(q)\:C\:\frac{1}{2}\:\frac{I_1}{I_Z}
\eeq
where 
\bee
I_1=\frac{1}{(4\pi)^{d/2}}\:
\frac{d-1}{d\,\left( 1 + 2\,\kappa  \right)\Gamma(d/2)}\: .
\eeq
Plugging in the value (\ref{IZ}) for $I_Z(\kappa)$, leads directly to Eq.\ (\ref{constC}).

Before one can calculate $\Gamma^{(GZZ)}_\mu(0;q,q)$, the three-gluon vertex is needed. According to Eqs.\ (\ref{ZZZDSEk0}) and (\ref{ZZZk0}), we can find the integral $I_3$ to yield
\bee
\label{I3}
I_3&=&\frac{(p^2)^{\alpha_G-\alpha_Z}}{d-1}\left(\Xi_1(1+2\kappa,1+\kappa)-\Xi_3(2+2\kappa,1+\kappa)/p^2\right)\nn\\
&=&\frac{1}{2(4\pi)^{d/2}(d-1)}\frac{\Gamma(\frac{d}{2} - 2\kappa )\Gamma(\frac{d}{2} - \kappa )\Gamma(2 - \frac{d}{2} + 3\kappa )}{\Gamma(d - 3\kappa )\Gamma(1 + \kappa )\Gamma(2 + 2\kappa )}
\: .
\eeq
Using this expression for the three-gluon vertex in Eq.\ (\ref{GGZDSE2ndk0}), we find that \bee
\label{I2}
\lim_{q\rarr 0}\Gamma^{(GZZ)}_\mu(0;q,q)&=&-ig^2A^2B^4I_3q_\mu\frac{N_c^2}{2}\int d^d\ellbar\:\ell\cdot q\left(1-(\hat p\cdot\hat\ell)^2\right)D_Z^2(\ell)D_G(\ell-p)(p^2)^{-(\alpha_G-\alpha_Z)} \nn\\
&=& -igq_\mu\:\frac{1}{2}\:\frac{I_3}{I_Z^2}\left(\Xi_1(d/2-\kappa,1+\kappa)-\Xi_3(d/2-\kappa+1,1+\kappa)/q^2
\right)\nn\\
&=& -\Gamma_\mu^{(0)}(q)\:\frac{1}{2}\:\frac{I_2I_3}{I_Z^2}
\eeq
where
\bee
I_2=\frac{1}{(4\pi)^{d/2}}\:\frac{d-1}{(d-2\kappa)\Gamma(d/2+1)}\; .
\eeq
Altogether, the infrared gluon limit of the ghost-gluon vertex, defined by Eq.\ (\ref{Cdef}) yields,
\bee
\label{constCfin}
C&=&1+\frac{1}{2}C\frac{I_1}{I_Z}-\frac{1}{2}\frac{I_2I_3}{I_Z^2}\nn\\
&=&2\left\{ 4^{\kappa }\left(d-1 \right) d\left( \Gamma(1 + \frac{d}{2}) - \kappa\: \Gamma(\frac{d}{2}) \right) 
     \Gamma(d - 3\kappa )\Gamma(\frac{d}{2} - \kappa ){\Gamma(-\kappa )}^2\Gamma(\frac{1}{2} + \kappa ) \Gamma(1 + 2\kappa )\right.\nn\\
&&\left.\left.\Gamma(2 + 2\kappa ) - 
    \sqrt{\pi }\:\Gamma(\frac{d}{2} - 2\kappa )^3\Gamma(1 + \frac{d}{2} + \kappa )^2\Gamma(2 - \frac{d}{2} + 3\kappa )\right\}\right/\nn\\
&&\left\{\left(d-1 \right) \,\left( d - 2\,\kappa  \right) \,\Gamma(d - 3\,\kappa )\,\Gamma(-\kappa )\,\Gamma(1 + 2\,\kappa )^2\right.\nn\\
&&\left.  \left( 4^{1 + \kappa }\Gamma(1 + \frac{d}{2})\Gamma(\frac{d}{2} - \kappa )\Gamma(-\kappa )
     \Gamma(\frac{3}{2} + \kappa ) + {\sqrt{\pi }}\Gamma(\frac{d}{2} - 2\kappa )\Gamma(1 + \frac{d}{2} + \kappa ) \right)\right\}
\eeq

As can be checked, this leads to the numerical values of $C$ given by Eq.\ (\ref{Cresult}) for the various solutions of $\kappa$.

\section*{Appendix B: The DSE for the three-gluon vertex}
\renewcommand{\theequation}{B\arabic{equation}}
\setcounter{equation}{0}  

The Dyson-Schwinger equation is derived from the generating functional $Z$ of the theory,
\begin{equation}
Z[j,\bar{\sigma},\sigma] =
\int\mathcal{D}[Ac\bar{c}] \exp \left[-\int d^dx \mathcal{L}+ \int d^dx \left(j_\mu^a A_\mu^a + \bar{\sigma}^a c^a + \bar{c}^a  \sigma^a \right)\right] \; ,
\end{equation}
where $\mathcal{L}$ is given by
\begin{equation}
\mathcal{L}(x) = \int d^d x' A^a_\mu ({ x} ) \omega ({ x}, { x}') A^a_\mu ({ x}') - \int d^dx' \bar{c}^a(x)(-\partial \cdot D[A])^{ab}(x,x') c^b(x') \; ,
\end{equation}
see Eq.\ (\ref{ZCoulomb}), with $\lambda=0$.

To derive the DSE we observe that
\begin{equation}
0 = \int \mathcal{D}[c\bar{c}A] \, \frac{\delta}{\delta
  A_\mu^c (u)} e^{-\int A \omega A + \int
  \bar{c}(-\partial\cdot D[A]) c + \int (jA + \bar{\sigma}c + \bar{c}\sigma)} \; ,
\end{equation}
as the integral can be turned into a surface integral over the Gribov horizon where the Faddeev-Popov determinant vanishes. We perform the derivative and replace emerging fields by derivative operators with respect to their sources in order to recover the generating functional $Z$. It is replaced according to $Z= e^W$, thus introducing the generating functional $W[j,\bar{\sigma},\sigma]$ of the connected Green functions:
\begin{equation}\label{first gluon}
\left[ -2 \int d^dx \; \omega(u,x) \; \frac{\delta}{\delta
    j_\mu^c(x)} + \int d^d[xx'] \frac{\delta^ R}{\delta \sigma^a
    (x)} (\Gamma_\mu^{0,c})^{ab} (u;x,x')
    \frac{\delta^L}{\delta\bar{\sigma}^b (x')} + j_\mu^c(u)\right] e^W = 0
\end{equation}

Here we use
\begin{equation}
\Gamma^{0,a}_\mu(x;y,z) = - \frac{\delta (-\partial\cdot D[A])(y,z)}{\delta A_\mu^a(x)}
\end{equation}
and
\begin{equation}
\frac{\delta^L}{\delta\bar{\sigma}} = \mbox{derivative acts from the left} ;\;\;\; \frac{\delta^ R}{\delta\sigma} = \mbox{derivative acts from the right} \; .
\end{equation}

The derivative in Eq.\ (\ref{first gluon}) corresponds to a first gluon of the three-gluon vertex. For the other two gluons we perform two further derivatives,
\begin{equation}\label{sec and third gluon}
\frac{\delta}{\delta j_\nu^d(v)} \;\;\mbox{and}\;\; \frac{\delta}{\delta j_\rho^e(w)}
\end{equation}
on Eq.\ (\ref{first gluon}). Setting $j=\sigma=\bar{\sigma}=0$ in Eq.\ (\ref{first gluon}) and in the equations obtained after each of the derivatives 
in Eq.\ (\ref{sec and third gluon}) yields the DSE for the connected one-, two- and three-point gluon Green functions respectively. Plugging the one- and two-point DSE into the three-point DSE we get
\begin{equation}
\begin{split}
 &\left. -2 \int d^dx \;\omega(u,x) \frac{\delta^3 W}{\delta j_\rho^e(w)
  \delta j_\nu^d (v) \delta j_\mu^c(x)} \right|_{J=0}\\
 &\left. + \int d^d[xx']
  (\Gamma_\mu^{0,c})^{ab} (u;x,x') \frac{\delta^4 W}{\delta
  j_\rho^e(w) \delta
  j_\nu^d(v) \delta \bar{\sigma}^b(x') \delta \sigma^a
  (x)}\right|_{J=0} = 0,
\end{split}
\end{equation}
where $J$ denotes all the sources collectively. After the application of
\begin{equation}
\int d^dz \;\omega^{-1}(y,z) \omega(z,x) = \delta^d(y-x)
\end{equation}
we can identify the gluon propagator $D_{\mu\nu}^{ab}$ on the RHS, cf. Eq.\ (37) in \cite{FeuRei04}. Introducing
\begin{equation}
\begin{split}
W_{\mu\nu\rho}^{abc}(x,y,z) :=& \left.\frac{\delta^3 W}{\delta j_\mu^a(x)
  \delta j_\nu^b (y) \delta j_\rho^c(z)}\right|_{J=0} \\
W_{\mu\nu}^{abcd}(w,x,y,z) :=& \left.\frac{\delta^4 W}{\delta
  j_\mu^a(w) \delta
  j_\nu^b(x) \delta \bar{\sigma}^c(y) \delta \sigma^d
  (z)}\right|_{J=0},
\end{split}
\end{equation}
provides as an intermediate result for the DSE for the connected three-gluon Green function:
\begin{equation}\label{conn dse}
 W_{\mu\nu\rho}^{abc}(x,y,z) = \int d^d[uvw] D_{\rho\kappa}^{cd}(z,u)
(\Gamma_\kappa^{0,d})^{ef} (u;v,w) W_{\mu\nu}^{abfe}(x,y,w,v)
\end{equation}

To derive a DSE for the proper three-gluon vertex we decompose the connected Green functions into the proper ones by using the generating functional $\Gamma$ of the proper Green functions which is the functional Legendre transform of $W$ defined by
\begin{equation}\label{leg}
\Gamma[A,\bar{c},c] := - W[j,\sigma,\bar{\sigma}] + \int d^dx
\;\left(A_\mu^a(x)j_\mu^a(x) + \bar{\sigma}^a(x)
c^a(x) + \bar{c}^a(x) \sigma^a(x)\right) 
\end{equation}
where the sources on the right side are chosen to fulfil
\begin{align}\label{leg1}
\frac{\delta W}{\delta j_\mu^a(x)} &= A_\mu^a(x) &
\frac{\delta W}{\delta \sigma^a(x)} &= \bar{c}^a(x) &
\frac{\delta W}{\delta \bar{\sigma}^a(x)} &= c^a(x)\; .
\end{align}

Therefrom we derive the relations
\begin{align}\label{leg2}
\frac{\delta \Gamma}{\delta A_\mu^a(x)} &= j_\mu^a(x) &
\frac{\delta \Gamma}{\delta \bar{c}^a(x)} &= \sigma^a(x) &
\frac{\delta \Gamma}{\delta c^a(x)} &= \bar{\sigma}^a(x) &
\end{align}
and the inversion relation
\begin{equation}\label{inversion}
\int d^dz \frac{\delta^2 W}{\delta j_\nu^b(y) \delta j_\rho^c(z)}
\frac{\delta^2 \Gamma}{\delta A_\rho^c(z) \delta A_\mu^a(x)} =
\int d^dz \frac{\delta A_\rho^c(z)}{\delta j_\nu^b(y)}
\frac{\delta j_\mu^a(x)}{\delta A_\rho^c(z)} = \delta^{ab} 
\delta_{\mu\nu} \delta^d(x-y) ,
\end{equation}
which is the starting point for the decomposition. We apply
\begin{equation}
\frac{\delta}{\delta j_\rho^d(u)}
\end{equation}
to Eq.\ (\ref{inversion}), use the chain rule and further apply Eq.\ (\ref{leg1}, \ref{leg2}, \ref{inversion}). With the definitions
\begin{equation}
 D_{\mu\nu}^{ab}(x,y) := \left.\frac{\delta^2 W}{\delta j_\mu^a(x) \delta
 j_\nu^b(y)}\right|_{J=0} \;\; \mbox{and} \qquad
\Gamma_{\mu\nu\rho}^{abc} (x,y,z) := \left.\frac{\delta^3 \Gamma}{\delta
 A_\mu^a(x) \delta A_\nu^b(y) \delta A_\rho^c(z)}\right|_{J=0}
\end{equation}
the decomposition of the connected three-gluon Green function reads
\begin{equation}\label{decomp 3gl}
W_{\mu\nu\rho}^{abc}(x,y,z) = - \int d^d[uvw]
\Gamma_{\kappa\lambda\sigma}^{def}(u,v,w) D_{\kappa\mu}^{da}(u,x)
D_{\lambda\nu}^{eb}(v,y) D_{\sigma\rho}^{fc}(w,z) \; ,
\end{equation}
which simply means cutting off the external propagators. This is different in the decomposition of the two-ghost-two-gluon vertex. The starting point is an inversion relation similar to Eq.\ (\ref{inversion}) that also is derived from the Legendre transformation:
\begin{equation}\label{inversion2}
\int d^dz \frac{\delta^2 W}{\delta \bar{\sigma}^a(x) \delta
  \sigma^c(z)} \frac{\delta^2 \Gamma}{\delta \bar{c}^c(z) \delta
  c^b(y)} = \int d^dz \frac{\delta \bar{\sigma}^b(y)}{\delta
  \bar{c}^c (z)}
  \frac{\delta\bar{c}^c(z)}{\delta\bar{\sigma}^a(x)} =
  \delta^{ab} \delta^d (x-y)
\end{equation}

To actually perform the decomposition it is useful to view the second derivatives as matrices where the colour, Lorentz and coordinate indices are matrix indices and the integration is the matrix multiplication. So we write Eq.\ (\ref{inversion2}) as
\begin{equation}\label{inversion matrixnotation}
\frac{\delta^2 W}{\delta \bar{\sigma} \delta\sigma} = \left[ \frac{\delta^2
  \Gamma}{\delta \bar{c} \delta c} \right] ^{-1} .
\end{equation}

From $\frac{\delta}{\delta t}(AA^{-1}) = 0$, where $A$ is a matrix, we obtain
\begin{equation}
\frac{\delta A^{-1}}{\delta t} = - A^{-1} \frac{\delta A}{\delta t} A^{-1} \; .
\end{equation}

We use this in taking the derivative of Eq.\ (\ref{inversion matrixnotation}) with respect to the gluon source:
\begin{equation}\label{3rdderivative}
\frac{\delta^3 W}{\delta j_\nu^b(x) \delta \bar{\sigma} \delta \sigma}
= - \left[ \frac{\delta^2
  \Gamma}{\delta \bar{c} \delta c} \right]^{-1}  \frac{\delta^3
  \Gamma}{\delta j_\nu^b(x) \delta \bar{c} \delta c} \left[\frac{\delta^2
  \Gamma}{\delta \bar{c} \delta c} \right] ^{-1}
\end{equation}

We apply Eq.\ (\ref{inversion matrixnotation}) and Eq.\ (\ref{leg1}) to this and take a further derivative with respect to the gluon source. Together with Eq.\ (\ref{3rdderivative}, \ref{decomp 3gl}) and making frequent use of the techniques just developed we obtain the decomposition of the connected two-gluon-two-ghost Green function. With the definitions

\begin{equation}
\begin{split}
&D_G^{ab}(x,y) := \left.\frac{\delta^2 W}{\delta \bar{\sigma}^a(x) \delta
 \sigma^b(y)}\right|_{J=0}  ,\qquad
\Gamma_\mu^{abc} (x,y,z) := \left.\frac{\delta^3 \Gamma}{\delta
 A_\mu^a(x) \delta \bar{c}^b(y) \delta c^c(z)}\right|_{J=0} \; ,\\
&\Gamma_{\mu\nu}^{abcd} (w,x,y,z) := \left.\frac{\delta^4 \Gamma}{\delta
 A_\mu^a(w) \delta A_\nu^b(x)\delta \bar{c}^c(y) \delta c^d(z)}\right|_{J=0}
\end{split}
\end{equation}
and switching back to the index notation it reads
\begin{equation}\label{4pt_decomp}
\begin{split}
&W_{\mu\nu}^{abcd}(w,x,y,z) =\\
&\int d^d[pqrstu] D_G^{ed}(s,z) D^{bf}_{\nu\rho}(x,t)
D_{\mu\kappa}^{ah}(w,p) D_G^{ci}(y,q) D_G^{jg}(r,u)\Gamma_\kappa^{hij}(p,q,r) \Gamma_\rho^{fge}(t,u,s)
\\
& + \int d^d[pqrstu] D_G^{ed}(s,z) D_G^{cf}(y,t)
D_{\mu\kappa}^{ah}(w,p) D_{\nu\lambda}^{bi}(x,q) D_{\rho\sigma}^{gj}(u,r)\Gamma_{\kappa\lambda\sigma}^{hij}(p,q,r) \Gamma_\rho^{gfe}(u,t,s)
\\
& + \int d^d [pqrstu] D_G^{ed}(s,z) D_{\mu\rho}^{af}(w,t)
D_{\nu\kappa}^{bh}(x,p) D_G^{ci}(y,q) D_G^{jg}(r,u)\Gamma_\kappa^{hij}(p,q,r) \Gamma_\rho^{fge}(t,u,s)
\\
&- \int d^d[stuv] D_G^{ed}(s,z) D_{\nu\rho}^{bf}(x,t)
D_{\mu\kappa}^{ag}(w,u) D_G^{ch}(y,v) \Gamma_{\kappa\rho}^{gfhe}(u,t,v,s) \; .
\end{split}
\end{equation}

Plugging Eq.\ (\ref{4pt_decomp}) and (\ref{decomp 3gl}) into the DSE for the connected functions, Eq.\ (\ref{conn dse}), and solving for the proper three-gluon vertex yields the DSE for it. It contains, however, a term that seems to be one-particle reducible. This comes about as we have not yet taken into account the DSE for the lower proper correlation functions. The two-gluon DSE is the equation obtained after the first of the two derivatives in Eq.\ (\ref{sec and third gluon}). Treating it the same way as the three-gluon DSE till the point of one-particle irreducibility gives
\begin{equation}
\int d^d[v_1v_2w_1w_2] (\Gamma_\mu^{0,a})^{cd} (x;v_1,v_2)
D_G^{de}(v_2,w_1) D_G^{fc}(w_2,v_1)
\Gamma_\nu^{bef}(y,w_1,w_2) = 0 \; ,
\end{equation}
which actually is found to be a part of the improper diagram in the three-gluon DSE that consequently vanishes.

We have thus arrived at the Dyson-Schwinger equation for the three-gluon vertex:
\begin{multline}\label{dse}
\Gamma_{\mu\nu\rho}^{abc}(x,y,z) = \\
\begin{aligned}
-2 \int d^d[u_1u_2v_1v_2w_1w_2] & (\Gamma_\rho^{0,c})^{de}(z;u_1,u_2)
  D_G^{fd}(w_2,u_1) D_G^{eg}(u_2,v_1) D_G^{hi}(v_2,w_1) \cdot
\\
\cdot & \Gamma_\mu^{agh}(x,v_1,v_2) \Gamma_\nu^{bif}(y,w_1,w_2)
\\
+ \int d^d[u_1u_2v_1v_2] & (\Gamma_\rho^{0,c})^{de}(z;u_1,u_2)
  D_G^{fd}(v_2,u_1) D_G^{eg}(u_2,v_1) \Gamma^{abgf}_{\mu\nu}(x,y,v_1,v_2)
\end{aligned}
\end{multline}


\begin{thebibliography}{99}


\bibitem{AlkSme00}
  R.~Alkofer and L.~von Smekal,
  Phys.\ Rept.\  {\bf 353} (2001) 281
  [arXiv:hep-ph/0007355];
  C.~S.~Fischer,
  arXiv:hep-ph/0605173.


\bibitem{AlkFisLla04}
  R.~Alkofer, C.~S.~Fischer and F.~J.~Llanes-Estrada,
  Phys.\ Lett.\ B {\bf 611} (2005) 279
  [arXiv:hep-th/0412330].

\bibitem{ChrLee80}
  N.~H.~Christ and T.~D.~Lee,
  Phys.\ Rev.\ D {\bf 22} (1980) 939
  [Phys.\ Scripta {\bf 23} (1981) 970].

\bibitem{FeuRei04}
  C.~Feuchter and H.~Reinhardt,
  Phys.\ Rev.\ D {\bf 70} (2004) 105021
  [arXiv:hep-th/0408236].

\bibitem{ReiFeu04}
  H.~Reinhardt and C.~Feuchter,
  Phys.\ Rev.\ D {\bf 71} (2005) 105002
  [arXiv:hep-th/0408237].

\bibitem{Szc}
A.~P.~Szczepaniak and E.~S.~Swanson,
  Phys.\ Rev.\ D {\bf 65} (2002) 025012
  [arXiv:hep-ph/0107078],
  A.~P.~Szczepaniak,
  Phys.\ Rev.\ D {\bf 69} (2004) 074031
  [arXiv:hep-ph/0306030].

\bibitem{LanReiGat01}
  K.~Langfeld, H.~Reinhardt and J.~Gattnar,
  Nucl.\ Phys.\ B {\bf 621} (2002) 131
  [arXiv:hep-ph/0107141],
J.~Gattnar, K.~Langfeld and H.~Reinhardt,
  Phys.\ Rev.\ Lett.\  {\bf 93} (2004) 061601
  [arXiv:hep-lat/0403011].


\bibitem{CucMenMih04}
  A.~Cucchieri, T.~Mendes and A.~Mihara,
  JHEP {\bf 0412} (2004) 012
  [arXiv:hep-lat/0408034].

\bibitem{Ste+05}
  A.~Sternbeck, E.~M.~Ilgenfritz, M.~Muller-Preussker and A.~Schiller,
  Nucl.\ Phys.\ Proc.\ Suppl.\  {\bf 153} (2006) 185
  [arXiv:hep-lat/0511053].

\bibitem{LanMoy04}
  K.~Langfeld and L.~Moyaerts,
  Phys.\ Rev.\ D {\bf 70} (2004) 074507
  [arXiv:hep-lat/0406024].

\bibitem{Zwa02}
  D.~Zwanziger,
  Phys.\ Rev.\ D {\bf 65} (2002) 094039
  [arXiv:hep-th/0109224].

\bibitem{LerSme02}
  C.~Lerche and L.~von Smekal,
  Phys.\ Rev.\ D {\bf 65} (2002) 125006
  [arXiv:hep-ph/0202194].

\bibitem{Sch+05}
  W.~Schleifenbaum, A.~Maas, J.~Wambach and R.~Alkofer,
  Phys.\ Rev.\ D {\bf 72} (2005) 014017
  [arXiv:hep-ph/0411052].

\bibitem{Zwa04}
  D.~Zwanziger,
  Phys.\ Rev.\ D {\bf 69} (2004) 016002
  [arXiv:hep-ph/0303028].

\bibitem{Zwa03a}
  D.~Zwanziger,
  Phys.\ Rev.\ D {\bf 70} (2004) 094034
  [arXiv:hep-ph/0312254].

\bibitem{Maas}
T.\ Appelquist and R.\ D.\ Pisarski,
Phys.\ Rev.\ D {\bf 23}, 2305 (1981);
A.\ Maas,
Mod.\ Phys.\ Lett.\ A {\bf 20}, 1797 (2005)
[arXiv:hep-ph/0506066].


\bibitem{Zwa91}
  D.~Zwanziger,
  Nucl.\ Phys.\ B {\bf 364} (1991) 127.

\bibitem{EppReiSch06}
  D.\ Epple et al., in preparation.

\bibitem{dipl_Schleifenbaum}
W.\ Schleifenbaum, 
diploma thesis, Technische Universit\"at Darmstadt, 2004.

\bibitem{Tay71}
  J.~C.~Taylor,
  Nucl.\ Phys.\ B {\bf 33} (1971) 436.

\bibitem{FisZwa05}
  C.~S.~Fischer and D.~Zwanziger,
  Phys.\ Rev.\ D {\bf 72} (2005) 054005
  [arXiv:hep-ph/0504244].

\bibitem{dipl_Leder}
M.~Leder, 
diploma thesis, Universit\"at T\"ubingen, 2006.

\bibitem{Zwa03b}
  D.~Zwanziger,
  Phys.\ Rev.\ Lett.\  {\bf 90} (2003) 102001
  [arXiv:hep-lat/0209105].


\end{thebibliography}
\end{document}